\documentclass[prl,aps,reprint,
 amsmath,amssymb,superscriptaddress]{revtex4-1}
\usepackage[utf8]{inputenc}
\usepackage{graphicx}
\usepackage{siunitx}
\usepackage{hyperref}

\begin{document}
\title{High-throughput and long-term observation of compartmentalized biochemical oscillators}

\author{Koshi Hasatani}
\affiliation{LIMMS/CNRS-IIS, University of Tokyo, Komaba 4-6-2 Meguro-ku, Tokyo, Japan}

\author{Mathieu Leocmach}
\affiliation{Institut Lumière Matière, UMR5306 Université Claude Bernard Lyon 1 - CNRS, Université de Lyon, 69622 Villeurbanne, France}

\author{Anthony J. Genot}
\affiliation{LIMMS/CNRS-IIS, University of Tokyo, Komaba 4-6-2 Meguro-ku, Tokyo, Japan}

\author{André Estévez-Torres}
\affiliation{Laboratoire de photonique et de nanostructures, CNRS, route de Nozay, 91460 Marcoussis, France}

\author{Teruo Fujii}
\affiliation{LIMMS/CNRS-IIS, University of Tokyo, Komaba 4-6-2 Meguro-ku, Tokyo, Japan}

\author{Yannick Rondelez}
\affiliation{LIMMS/CNRS-IIS, University of Tokyo, Komaba 4-6-2 Meguro-ku, Tokyo, Japan}
\email{rondelez@iis.u-tokyo.ac.jp}

\begin{abstract}
We report the splitting of an oscillating DNA circuit into $\sim 700$ droplets with picoliter volumes. Upon incubation at constant temperature, the droplets display sustained oscillations that can be observed for more than a day. Superimposed to the bulk behaviour, we find two intriguing new phenoma -- slow desynchronization between the compartments and kinematic spatial waves -- and investigate their possible origins. This approach provides a route to study the influence of small volume effects in biology, and paves the way to technological applications of compartmentalized molecular programs controlling complex dynamics.
\end{abstract}
\maketitle

Recent progresses in molecular engineering have allowed the construction of synthetic biochemical systems that display  subtle dynamics such as oscillations or multistability~\cite{Kim2006,Montagne2011,Kim2011,Padirac2012}. The building of such molecular programs from scratch yields a unique glimpse of the challenges and locks that evolution had to overcome in its drift towards more and more complex life forms. Moreover, the inherent biocompatibility of these chemical circuits supports their use to monitor and control biological processes. 

	Compartmentalization of these man-made dynamic systems would offer tantalizing additional possibilities. Cell-sized compartments ($\sim 1-\SI{100}{\micro\metre}$) may provide a model of biotic or prebiotic organization~\cite{Szostak2001}. For example, stochastic effects arising from low numbers of molecules in small spaces are very important in biology~\cite{Tkacik2009,Perkins2009}, but hardly explored outside of living systems. Synchronization of many autonomous elements using quorum sensing provides another example~\cite{Taylor2009}. 
	
	Beside these basic biological motivations, micro-compartmentalization is also the obvious way forward for the exploration of man-made molecular circuits. First, it permits the running of hundreds of circuits using volumes that would have yielded a single experiment otherwise. Such high-throughput will be required to tune the many parameters controlling the behavior of molecular assemblies. Even in the case where all the compartments possess an identical ``program'', compartmentalization may yield new analytical concepts such as digital PCR~\cite{Vogelstein1999}. Second, splitting - and later establishing controlled connections between - spatially distributed molecular circuits may unleash the computing power of molecular programs~\cite{Toiya2008} by removing cross-talks and allowing the reuse of modules.
	
 	Yet, compartmentalization and long-term monitoring of biochemical reactions in micro-compartments has been difficult to achieve. Indeed, the reactions involved must run sustainably in closed systems and the potentially detrimental effects of large surface/volume ratios and leaks need to be tightly controlled.
 	
\begin{figure}
\includegraphics{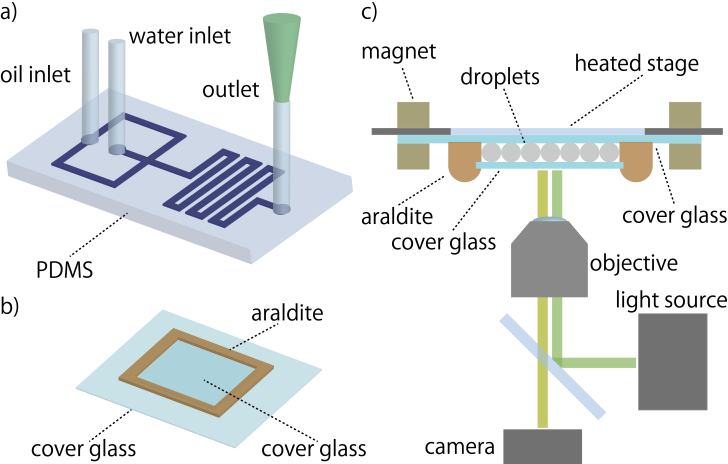}
\caption{a) Water-in-oil droplets are generated inside a PDMS chip with a flow-focusing junction. b) Droplets are then transferred to a glass chamber sealed with araldite for incubation and observation. c) The chamber is placed under a heated stage and observed by fluorescence microscopy.}
\label{fig:setup}
\end{figure}
 	
 	While biochemical systems with a simple dynamic such as qPCR were monitored in droplets~\cite{Guo2012}, out-of-equilibrium biochemical circuits pose distinct and additional challenges. PCR is driven sequentially by an external operator, uses only a single enzyme and lasts rarely more than 45 minutes. By contrast, our dynamic systems display oscillatory behaviour, are fully autonomous, require 3 enzymes, and last for days.
 	
	More importantly, sustainability of oscillations depends on a much more delicate balance between reagents, buffer and temperature than PCR. Most biochemical assays typically consist in a simple relaxation toward a stable steady state, but oscillations arise from the destabilization of all steady states, which requires tight control over reaction parameters. 
	
	Here we report the compartmentalization and day-long tracking of ~700 oscillators into picoliter droplets. This represents an increase of throughput by two orders of magnitudes and a decrease of volume by 4 orders of magnitudes compared to the literature on synthetic DNA systems1. We show that with an experimental setup minimizing evaporation, droplets act as independent chemical containers that satisfyingly reproduce bulk conditions. Moreover, compartmentalization also reveals two striking phenomena that may remain hidden in bulk: slow desynchronization and kinematic spatial waves. 
	
	We use here a recently reported synthetic biochemical oscillator~\cite{Fujii2013}. While simple to prepare and well characterized, it involves subtle enzymatic kinetics, which should put a stringent test on the use of droplets as compartments. The biochemical system reproduces the ecological predator-prey mechanism: molecular preys catalyse their own replication, but also serve as fuel for the replication of their molecular predators. Both prey and predators are continuously degraded by an exonuclease. This system results in robust oscillations, which are monitored via the fluorescence intensity of a dye bound to the DNA template encoding the circuit~\cite{Padirac2012a}.
	
	Droplets were generated by a flow-focusing junction fabricated in a PDMS chip. For each run, we generated thousands of droplets from the same reactive mix. The droplets were then transferred to a chamber, which was formed between two glass slides sealed by araldite. The chamber was kept at constant temperature (\SI{45.5}{\celsius}) by placing it under the heated glass plate of a microscope stage equipped with a temperature controller (Fig.~\ref{fig:setup}). A field of view contained about 1000 droplets (movies in ESI). While the oscillations lasted for 2 days, we restricted the analysis to one day in order to increase the number of trajectories successfully tracked~\cite{Crocker1996,Besseling2009,Leocmach2013}.

\begin{figure*}
\includegraphics[width=\textwidth]{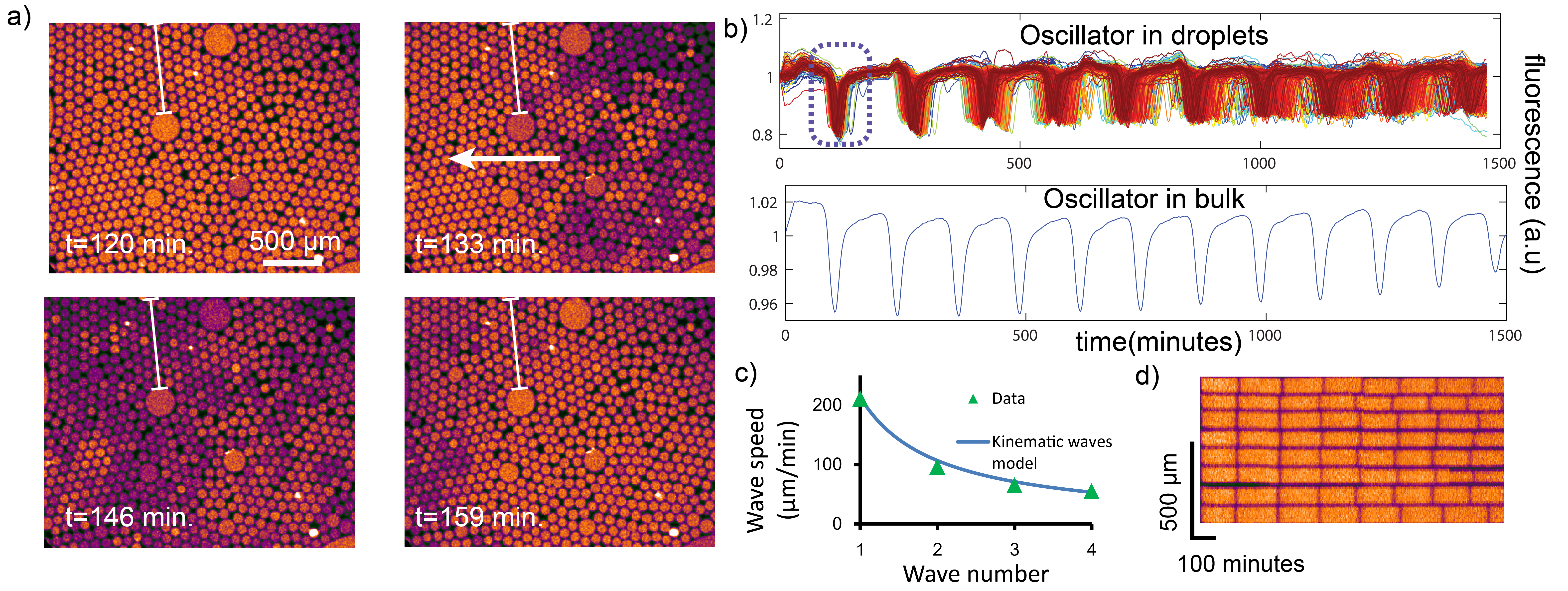}
\caption{a) Microscopy images showing the first kinematic spatial wave. The arrow indicates the travelling direction (colours reflect the intensity of fluorescence).  The temperature is set at \SI{45.5}{\celsius}. b) Top. Superimposed fluorescence traces of all successfully tracked droplets (690 traces). The dashed box indicates the peak corresponding to the wave shown in a). Bottom, bulk fluorescence of the oscillator (set at the same temperature) measured in a \SI{20}{\micro\litre} tube in a qPCR machine. c) Wave speed as a function of the wave number, fitted with a kinematic wave model (S11, ESI). d) Kymograph showing the desynchronization of a column of droplets lying on the same initial wavefront (column indicated by a vertical line in a)). }
\label{fig:oscillations}
\end{figure*}

	Fig.~\ref{fig:oscillations}a shows a representative field of view and time traces for the fluorescence of droplets. Most droplets are monodisperse, with a diameter of $\sim\SI{100}{\micro\metre}$ and a dispersion of $\sim15\%$. This corresponds to a volume of $\sim 500$ picoliters, which is 4 orders of magnitude smaller than bulk reaction volume (\SI{20}{\micro\litre}). A few droplets are significantly larger, which may result from coalescence during the generation and transfer of droplets. The fluorescence of the majority of droplets oscillates, and the oscillation of each individual droplet is similar to the bulk control.
	
	The collective behaviour was more surprising: spatial waves move across the chamber while slow desynchronization ultimately sets each droplet’s fluorescence on a seemingly independent trajectory (Fig.~\ref{fig:oscillations}b). 
Since the contents of all droplets derive from the same mix, we expect them to be initially synchronized. Indeed, noting $T_i$ the time required to reach the $i^\text{th}$ peak (Fig.~\ref{fig:oscillations}a), we measure a standard deviation of 7 minutes for $T_1$. After 15 hours, the standard deviation of $T_7$ had increased to 35 minutes (25\% of the mean period). At this point no coherence can be visually detected in the droplet population. We tested if this desynchronization was specific to compartmentalization or also existed in bulk. In \SI{20}{\micro\litre} tubes, we observed a similar albeit less pronounced desynchronization (S8, ESI). This common desynchronization points to the influence of an external factor such as temperature on the period of oscillations. We characterised the sensitivity of oscillator to temperature in bulk (S9, ESI) and found that an increase of temperature of \SI{2}{\celsius} lengthens the period by 25\%.

Desynchronization is not spatially random, but manifests itself initially as travelling waves (movies  M1 and Fig.~\ref{fig:oscillations}a). One may misinterpret spatial waves as the result of diffusive transport of reagents. However, pseudo (also called kinematic) waves may appear when properties determining phases or periods vary spatially~\cite{Kopell1973}. In view of the temperature-sensitivity of the oscillator, a gradient of temperature could not only desynchronize the droplets, but also generates kinematic waves. 

	The kinematic wave model predicts in particular a wave speed inversely proportional to the wave number, which we observed experimentally (Fig.~\ref{fig:setup}c, S11 ESI). The kinematic model also implies a thermal gradient of \SI{0.3}{\kelvin\per\milli\metre}, which agrees with the measured gradient of \SI{0.15}{\kelvin\per\milli\metre} (S10, ESI).
	
	On the contrary, the experimentally observed speeds of the waves do not support a model where waves are caused by diffusion of DNA strands between droplets. This model would be consistent with a time for droplet-to-droplet diffusion of $\sim 5$ minutes (S11,ESI). 
	
This timescale is 2 orders of magnitude larger than the diffusion of a smaller molecule (fluorescein) between droplets stabilized by a more permeable surfactant (Span-80)~\cite{Bai2010}. The timescale is also incompatible with the possibility of digital PCR in droplets (which lasts ~30 minutes)~\cite{Pekin2011}.

\begin{figure}
\includegraphics{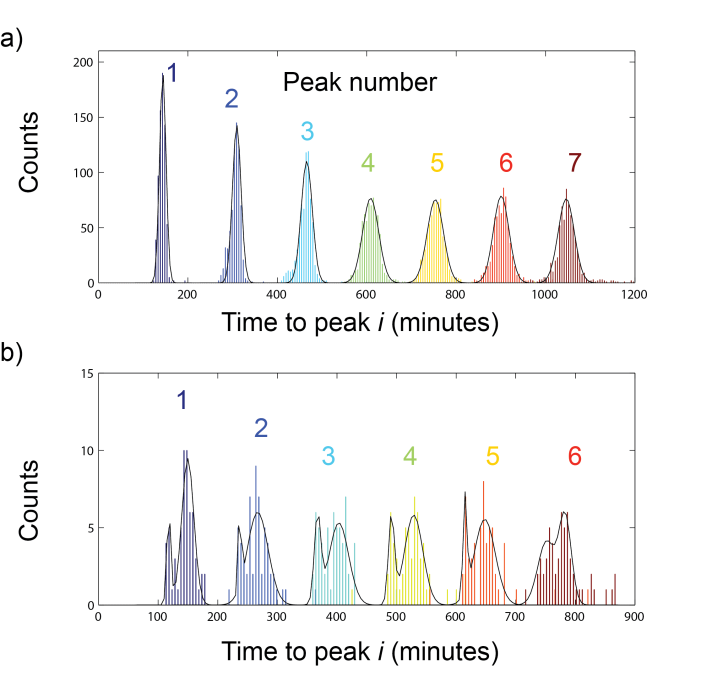}
\caption{Distribution of  $T_i$, the time it takes for the fluorescence of a droplet to reach its $i^\text{th}$  (extinction)  peak, measured from the beginning of the recording. a) Histogram of $T_i$ for the 690 traces from Fig.~\ref{fig:oscillations}b, for the first 7 peaks. b) Histogram of $T_i$ for 59 traces of a mixed-population experiment where the oscillations of one population are delayed compared to the other. The black line shows the best fit with a single (a) or two (b) Gaussian distributions to each peak.}
\label{fig:peaks}
\end{figure}

To further test that diffusion between droplets was negligible, we prepared two populations of oscillating droplets with a delay between them. One population was pre-incubated at reaction temperature (\SI{45.5}{\celsius}) for approximately half a period, thus allowing the oscillation to progress. The remaining population was kept at room temperature where reactions are much slower. The two populations were then mixed and incubated together on stage. If diffusion between droplets is negligible, the delay between the two populations of droplets should persist over time. The movie M2 shows the sustained presence of two populations. Fig.~\ref{fig:peaks}b shows the distribution of time-to-peak for this experiment. The initial delay is clearly observable from the shape of $T_1$, where the populations are easily distinguished. This bimodal distribution was maintained at least until $T_6$ despite the widening of both populations (overall standard deviation of 16 minutes at $T_1$ versus 51 minutes for $T_6$, compared to 7 and 29 minutes for T1 and $T_6$ (in Fig.~\ref{fig:peaks}a).

This population stability supports the functioning of each droplet as an independent reactor. The kinematic waves suggest that droplets are essentially identical but placed in a thermal gradient. Yet, some droplet-specific desynchronization is also visible in the movies (Fig.~\ref{fig:oscillations}d) and we cannot rule out that intrinsic factors such as stochasticity or partitioning noise play a role in the observed dynamic. Predator-prey oscillators are nonlinear and their autocatalytic loops may amplify minute differences or fluctuations between droplets. Noise-induced stochasticity is an important determinant of biological circuits dynamic.

	In conclusion, the picoliter droplet-based compartments used here satisfactorily reproduce the behaviours observed in bulk, in spite of reducing volumes by a factor of $10^4$.  Therefore this approach may open the high-throughput experimental exploration~\cite{Pompano2011} of the range of dynamics displayed by synthetic molecular programs when varying their control parameters. Another interesting direction would be to engineer a form of controlled molecular communication between the droplets in order to trigger and tune the onset of collective behaviours4.  A further shrinking in size would better match the dimensions of biological cells and provides a unique platform to study the impact of small molecule numbers on the dynamics of dissipative reaction networks. 
	
	Compartmentalized oscillators also offer an in vitro model to investigate pattern-formation in biology. Kinematic waves clearly illustrate that spatio-temporal patterns - ubiquitous in biological development- need not arise from a reaction-diffusion mechanism, but may also originate from a spatial gradient of parameters. Purposely tailoring spatial gradients of parameters will offer a route to create novel classes of patterns.
	
\begin{acknowledgments}
	This research is supported by the CNRS (France) and the JSPS (Japan). We thank V. Taly, H. Guillou and A. Padirac for advice as well as B.J. Kim and R. Ueno for using their thermal camera. During the course of our research, E. Winfree and F. Simmel kindly informed us of their ongoing work on the effect of compartmentalization and stochasticity in biochemical oscillators.
\end{acknowledgments}

\bibliographystyle{apsrev4-1}
\bibliography{oscillators}

\begin{thebibliography}{20}%
\makeatletter
\providecommand \@ifxundefined [1]{%
 \@ifx{#1\undefined}
}%
\providecommand \@ifnum [1]{%
 \ifnum #1\expandafter \@firstoftwo
 \else \expandafter \@secondoftwo
 \fi
}%
\providecommand \@ifx [1]{%
 \ifx #1\expandafter \@firstoftwo
 \else \expandafter \@secondoftwo
 \fi
}%
\providecommand \natexlab [1]{#1}%
\providecommand \enquote  [1]{``#1''}%
\providecommand \bibnamefont  [1]{#1}%
\providecommand \bibfnamefont [1]{#1}%
\providecommand \citenamefont [1]{#1}%
\providecommand \href@noop [0]{\@secondoftwo}%
\providecommand \href [0]{\begingroup \@sanitize@url \@href}%
\providecommand \@href[1]{\@@startlink{#1}\@@href}%
\providecommand \@@href[1]{\endgroup#1\@@endlink}%
\providecommand \@sanitize@url [0]{\catcode `\\12\catcode `\$12\catcode
  `\&12\catcode `\#12\catcode `\^12\catcode `\_12\catcode `\%12\relax}%
\providecommand \@@startlink[1]{}%
\providecommand \@@endlink[0]{}%
\providecommand \url  [0]{\begingroup\@sanitize@url \@url }%
\providecommand \@url [1]{\endgroup\@href {#1}{\urlprefix }}%
\providecommand \urlprefix  [0]{URL }%
\providecommand \Eprint [0]{\href }%
\providecommand \doibase [0]{http://dx.doi.org/}%
\providecommand \selectlanguage [0]{\@gobble}%
\providecommand \bibinfo  [0]{\@secondoftwo}%
\providecommand \bibfield  [0]{\@secondoftwo}%
\providecommand \translation [1]{[#1]}%
\providecommand \BibitemOpen [0]{}%
\providecommand \bibitemStop [0]{}%
\providecommand \bibitemNoStop [0]{.\EOS\space}%
\providecommand \EOS [0]{\spacefactor3000\relax}%
\providecommand \BibitemShut  [1]{\csname bibitem#1\endcsname}%
\let\auto@bib@innerbib\@empty
%</preamble>
\bibitem [{\citenamefont {Kim}\ \emph {et~al.}(2006)\citenamefont {Kim},
  \citenamefont {White},\ and\ \citenamefont {Winfree}}]{Kim2006}%
  \BibitemOpen
  \bibfield  {author} {\bibinfo {author} {\bibfnamefont {J.}~\bibnamefont
  {Kim}}, \bibinfo {author} {\bibfnamefont {K.~S.}\ \bibnamefont {White}}, \
  and\ \bibinfo {author} {\bibfnamefont {E.}~\bibnamefont {Winfree}},\ }\href
  {\doibase 10.1038/msb4100099} {\bibfield  {journal} {\bibinfo  {journal}
  {Mol. Syst. Biol.}\ }\textbf {\bibinfo {volume} {2}} (\bibinfo {year}
  {2006}),\ 10.1038/msb4100099}\BibitemShut {NoStop}%
\bibitem [{\citenamefont {Montagne}\ \emph {et~al.}(2011)\citenamefont
  {Montagne}, \citenamefont {Plasson}, \citenamefont {Sakai}, \citenamefont
  {Fujii},\ and\ \citenamefont {Rondelez}}]{Montagne2011}%
  \BibitemOpen
  \bibfield  {author} {\bibinfo {author} {\bibfnamefont {K.}~\bibnamefont
  {Montagne}}, \bibinfo {author} {\bibfnamefont {R.}~\bibnamefont {Plasson}},
  \bibinfo {author} {\bibfnamefont {Y.}~\bibnamefont {Sakai}}, \bibinfo
  {author} {\bibfnamefont {T.}~\bibnamefont {Fujii}}, \ and\ \bibinfo {author}
  {\bibfnamefont {Y.}~\bibnamefont {Rondelez}},\ }\href {\doibase
  10.1038/msb.2010.120} {\bibfield  {journal} {\bibinfo  {journal} {Mol. Syst.
  Biol.}\ }\textbf {\bibinfo {volume} {7}},\ \bibinfo {pages} {466} (\bibinfo
  {year} {2011})}\BibitemShut {NoStop}%
\bibitem [{\citenamefont {Kim}\ and\ \citenamefont {Winfree}(2011)}]{Kim2011}%
  \BibitemOpen
  \bibfield  {author} {\bibinfo {author} {\bibfnamefont {J.}~\bibnamefont
  {Kim}}\ and\ \bibinfo {author} {\bibfnamefont {E.}~\bibnamefont {Winfree}},\
  }\href {\doibase 10.1038/msb.2010.119} {\bibfield  {journal} {\bibinfo
  {journal} {Mol. Syst. Biol.}\ }\textbf {\bibinfo {volume} {7}},\ \bibinfo
  {pages} {465} (\bibinfo {year} {2011})}\BibitemShut {NoStop}%
\bibitem [{\citenamefont {Padirac}\ \emph
  {et~al.}(2012{\natexlab{a}})\citenamefont {Padirac}, \citenamefont {Fujii},\
  and\ \citenamefont {Rondelez}}]{Padirac2012}%
  \BibitemOpen
  \bibfield  {author} {\bibinfo {author} {\bibfnamefont {A.}~\bibnamefont
  {Padirac}}, \bibinfo {author} {\bibfnamefont {T.}~\bibnamefont {Fujii}}, \
  and\ \bibinfo {author} {\bibfnamefont {Y.}~\bibnamefont {Rondelez}},\ }\href
  {\doibase 10.1073/pnas.1212069109} {\bibfield  {journal} {\bibinfo  {journal}
  {Proc. Natl. Acad. Sci.}\ }\textbf {\bibinfo {volume} {109}},\ \bibinfo
  {pages} {E3212} (\bibinfo {year} {2012}{\natexlab{a}})}\BibitemShut {NoStop}%
\bibitem [{\citenamefont {Szostak}\ \emph {et~al.}(2001)\citenamefont
  {Szostak}, \citenamefont {Bartel},\ and\ \citenamefont
  {Luisi}}]{Szostak2001}%
  \BibitemOpen
  \bibfield  {author} {\bibinfo {author} {\bibfnamefont {J.~W.}\ \bibnamefont
  {Szostak}}, \bibinfo {author} {\bibfnamefont {D.~P.}\ \bibnamefont {Bartel}},
  \ and\ \bibinfo {author} {\bibfnamefont {P.~L.}\ \bibnamefont {Luisi}},\
  }\href {\doibase 10.1038/35053176} {\bibfield  {journal} {\bibinfo  {journal}
  {Nature}\ }\textbf {\bibinfo {volume} {409}},\ \bibinfo {pages} {387}
  (\bibinfo {year} {2001})}\BibitemShut {NoStop}%
\bibitem [{\citenamefont {Tka{\v{c}}ik}\ and\ \citenamefont
  {Bialek}(2009)}]{Tkacik2009}%
  \BibitemOpen
  \bibfield  {author} {\bibinfo {author} {\bibfnamefont {G.}~\bibnamefont
  {Tka{\v{c}}ik}}\ and\ \bibinfo {author} {\bibfnamefont {W.}~\bibnamefont
  {Bialek}},\ }\href {\doibase 10.1103/PhysRevE.79.051901} {\bibfield
  {journal} {\bibinfo  {journal} {Phys. Rev. E}\ }\textbf {\bibinfo {volume}
  {79}},\ \bibinfo {pages} {051901} (\bibinfo {year} {2009})}\BibitemShut
  {NoStop}%
\bibitem [{\citenamefont {Perkins}\ and\ \citenamefont
  {Swain}(2009)}]{Perkins2009}%
  \BibitemOpen
  \bibfield  {author} {\bibinfo {author} {\bibfnamefont {T.~J.}\ \bibnamefont
  {Perkins}}\ and\ \bibinfo {author} {\bibfnamefont {P.~S.}\ \bibnamefont
  {Swain}},\ }\href {\doibase 10.1038/msb.2009.83} {\bibfield  {journal}
  {\bibinfo  {journal} {Mol. Syst. Biol.}\ }\textbf {\bibinfo {volume} {5}}
  (\bibinfo {year} {2009}),\ 10.1038/msb.2009.83}\BibitemShut {NoStop}%
\bibitem [{\citenamefont {Taylor}\ \emph {et~al.}(2009)\citenamefont {Taylor},
  \citenamefont {Tinsley}, \citenamefont {Wang}, \citenamefont {Huang},\ and\
  \citenamefont {Showalter}}]{Taylor2009}%
  \BibitemOpen
  \bibfield  {author} {\bibinfo {author} {\bibfnamefont {A.~F.}\ \bibnamefont
  {Taylor}}, \bibinfo {author} {\bibfnamefont {M.~R.}\ \bibnamefont {Tinsley}},
  \bibinfo {author} {\bibfnamefont {F.}~\bibnamefont {Wang}}, \bibinfo {author}
  {\bibfnamefont {Z.}~\bibnamefont {Huang}}, \ and\ \bibinfo {author}
  {\bibfnamefont {K.}~\bibnamefont {Showalter}},\ }\href {\doibase
  10.1126/science.1166253} {\bibfield  {journal} {\bibinfo  {journal} {Science
  (80-. ).}\ }\textbf {\bibinfo {volume} {323}},\ \bibinfo {pages} {614}
  (\bibinfo {year} {2009})}\BibitemShut {NoStop}%
\bibitem [{\citenamefont {Vogelstein}\ and\ \citenamefont
  {Kinzler}(1999)}]{Vogelstein1999}%
  \BibitemOpen
  \bibfield  {author} {\bibinfo {author} {\bibfnamefont {B.}~\bibnamefont
  {Vogelstein}}\ and\ \bibinfo {author} {\bibfnamefont {K.~W.}\ \bibnamefont
  {Kinzler}},\ }\href {\doibase 10.1073/pnas.96.16.9236} {\bibfield  {journal}
  {\bibinfo  {journal} {Proc. Natl. Acad. Sci.}\ }\textbf {\bibinfo {volume}
  {96}},\ \bibinfo {pages} {9236} (\bibinfo {year} {1999})}\BibitemShut
  {NoStop}%
\bibitem [{\citenamefont {Toiya}\ \emph {et~al.}(2008)\citenamefont {Toiya},
  \citenamefont {Vanag},\ and\ \citenamefont {Epstein}}]{Toiya2008}%
  \BibitemOpen
  \bibfield  {author} {\bibinfo {author} {\bibfnamefont {M.}~\bibnamefont
  {Toiya}}, \bibinfo {author} {\bibfnamefont {V.~K.}\ \bibnamefont {Vanag}}, \
  and\ \bibinfo {author} {\bibfnamefont {I.~R.}\ \bibnamefont {Epstein}},\
  }\href {\doibase 10.1002/anie.200802339} {\bibfield  {journal} {\bibinfo
  {journal} {Angew. Chemie Int. Ed.}\ }\textbf {\bibinfo {volume} {47}},\
  \bibinfo {pages} {7753} (\bibinfo {year} {2008})}\BibitemShut {NoStop}%
\bibitem [{\citenamefont {Guo}\ \emph {et~al.}(2012)\citenamefont {Guo},
  \citenamefont {Rotem}, \citenamefont {Heyman},\ and\ \citenamefont
  {Weitz}}]{Guo2012}%
  \BibitemOpen
  \bibfield  {author} {\bibinfo {author} {\bibfnamefont {M.~T.}\ \bibnamefont
  {Guo}}, \bibinfo {author} {\bibfnamefont {A.}~\bibnamefont {Rotem}}, \bibinfo
  {author} {\bibfnamefont {J.~A.}\ \bibnamefont {Heyman}}, \ and\ \bibinfo
  {author} {\bibfnamefont {D.~A.}\ \bibnamefont {Weitz}},\ }\href {\doibase
  10.1039/c2lc21147e} {\bibfield  {journal} {\bibinfo  {journal} {Lab Chip}\
  }\textbf {\bibinfo {volume} {12}},\ \bibinfo {pages} {2146} (\bibinfo {year}
  {2012})}\BibitemShut {NoStop}%
\bibitem [{\citenamefont {Fujii}\ and\ \citenamefont
  {Rondelez}(2013)}]{Fujii2013}%
  \BibitemOpen
  \bibfield  {author} {\bibinfo {author} {\bibfnamefont {T.}~\bibnamefont
  {Fujii}}\ and\ \bibinfo {author} {\bibfnamefont {Y.}~\bibnamefont
  {Rondelez}},\ }\href {\doibase 10.1021/nn3043572} {\bibfield  {journal}
  {\bibinfo  {journal} {ACS Nano}\ }\textbf {\bibinfo {volume} {7}},\ \bibinfo
  {pages} {27} (\bibinfo {year} {2013})}\BibitemShut {NoStop}%
\bibitem [{\citenamefont {Padirac}\ \emph
  {et~al.}(2012{\natexlab{b}})\citenamefont {Padirac}, \citenamefont {Fujii},\
  and\ \citenamefont {Rondelez}}]{Padirac2012a}%
  \BibitemOpen
  \bibfield  {author} {\bibinfo {author} {\bibfnamefont {A.}~\bibnamefont
  {Padirac}}, \bibinfo {author} {\bibfnamefont {T.}~\bibnamefont {Fujii}}, \
  and\ \bibinfo {author} {\bibfnamefont {Y.}~\bibnamefont {Rondelez}},\ }\href
  {\doibase 10.1093/nar/gks621} {\bibfield  {journal} {\bibinfo  {journal}
  {Nucleic Acids Res.}\ }\textbf {\bibinfo {volume} {40}},\ \bibinfo {pages}
  {e118} (\bibinfo {year} {2012}{\natexlab{b}})}\BibitemShut {NoStop}%
\bibitem [{\citenamefont {Crocker}\ and\ \citenamefont
  {Grier}(1996)}]{Crocker1996}%
  \BibitemOpen
  \bibfield  {author} {\bibinfo {author} {\bibfnamefont {J.~C.}\ \bibnamefont
  {Crocker}}\ and\ \bibinfo {author} {\bibfnamefont {D.~G.}\ \bibnamefont
  {Grier}},\ }\href {\doibase 10.1006/jcis.1996.0217} {\bibfield  {journal}
  {\bibinfo  {journal} {J. Coll. Interf. Sci.}\ }\textbf {\bibinfo {volume}
  {179}},\ \bibinfo {pages} {298} (\bibinfo {year} {1996})}\BibitemShut
  {NoStop}%
\bibitem [{\citenamefont {Besseling}\ \emph {et~al.}(2009)\citenamefont
  {Besseling}, \citenamefont {Isa}, \citenamefont {Weeks},\ and\ \citenamefont
  {Poon}}]{Besseling2009}%
  \BibitemOpen
  \bibfield  {author} {\bibinfo {author} {\bibfnamefont {R.}~\bibnamefont
  {Besseling}}, \bibinfo {author} {\bibfnamefont {L.}~\bibnamefont {Isa}},
  \bibinfo {author} {\bibfnamefont {E.~R.}\ \bibnamefont {Weeks}}, \ and\
  \bibinfo {author} {\bibfnamefont {W.~C.}\ \bibnamefont {Poon}},\ }\href
  {\doibase 10.1016/j.cis.2008.09.008} {\bibfield  {journal} {\bibinfo
  {journal} {Adv. Colloid Interface Sci.}\ }\textbf {\bibinfo {volume} {146}},\
  \bibinfo {pages} {1} (\bibinfo {year} {2009})}\BibitemShut {NoStop}%
\bibitem [{\citenamefont {Leocmach}\ and\ \citenamefont
  {Tanaka}(2013)}]{Leocmach2013}%
  \BibitemOpen
  \bibfield  {author} {\bibinfo {author} {\bibfnamefont {M.}~\bibnamefont
  {Leocmach}}\ and\ \bibinfo {author} {\bibfnamefont {H.}~\bibnamefont
  {Tanaka}},\ }\href {\doibase 10.1039/c2sm27107a} {\bibfield  {journal}
  {\bibinfo  {journal} {Soft Matter}\ }\textbf {\bibinfo {volume} {9}},\
  \bibinfo {pages} {1447} (\bibinfo {year} {2013})},\ \Eprint
  {http://arxiv.org/abs/1301.7237} {arXiv:1301.7237} \BibitemShut {NoStop}%
\bibitem [{\citenamefont {Kopell}\ and\ \citenamefont
  {Howard}(1973)}]{Kopell1973}%
  \BibitemOpen
  \bibfield  {author} {\bibinfo {author} {\bibfnamefont {N.}~\bibnamefont
  {Kopell}}\ and\ \bibinfo {author} {\bibfnamefont {L.~N.}\ \bibnamefont
  {Howard}},\ }\href {\doibase 10.1126/science.180.4091.1171} {\bibfield
  {journal} {\bibinfo  {journal} {Science (80-. ).}\ }\textbf {\bibinfo
  {volume} {180}},\ \bibinfo {pages} {1171} (\bibinfo {year}
  {1973})}\BibitemShut {NoStop}%
\bibitem [{\citenamefont {Bai}\ \emph {et~al.}(2010)\citenamefont {Bai},
  \citenamefont {He}, \citenamefont {Liu}, \citenamefont {Patil}, \citenamefont
  {Bratton}, \citenamefont {Huebner}, \citenamefont {Hollfelder}, \citenamefont
  {Abell},\ and\ \citenamefont {Huck}}]{Bai2010}%
  \BibitemOpen
  \bibfield  {author} {\bibinfo {author} {\bibfnamefont {Y.}~\bibnamefont
  {Bai}}, \bibinfo {author} {\bibfnamefont {X.}~\bibnamefont {He}}, \bibinfo
  {author} {\bibfnamefont {D.}~\bibnamefont {Liu}}, \bibinfo {author}
  {\bibfnamefont {S.~N.}\ \bibnamefont {Patil}}, \bibinfo {author}
  {\bibfnamefont {D.}~\bibnamefont {Bratton}}, \bibinfo {author} {\bibfnamefont
  {A.}~\bibnamefont {Huebner}}, \bibinfo {author} {\bibfnamefont
  {F.}~\bibnamefont {Hollfelder}}, \bibinfo {author} {\bibfnamefont
  {C.}~\bibnamefont {Abell}}, \ and\ \bibinfo {author} {\bibfnamefont
  {W.~T.~S.}\ \bibnamefont {Huck}},\ }\href {\doibase 10.1039/b925133b}
  {\bibfield  {journal} {\bibinfo  {journal} {Lab Chip}\ }\textbf {\bibinfo
  {volume} {10}},\ \bibinfo {pages} {1281} (\bibinfo {year}
  {2010})}\BibitemShut {NoStop}%
\bibitem [{\citenamefont {Pekin}\ \emph {et~al.}(2011)\citenamefont {Pekin},
  \citenamefont {Skhiri}, \citenamefont {Baret}, \citenamefont {{Le Corre}},
  \citenamefont {Mazutis}, \citenamefont {{Ben Salem}}, \citenamefont {Millot},
  \citenamefont {{El Harrak}}, \citenamefont {Hutchison}, \citenamefont
  {Larson}, \citenamefont {Link}, \citenamefont {Laurent-Puig}, \citenamefont
  {Griffiths},\ and\ \citenamefont {Taly}}]{Pekin2011}%
  \BibitemOpen
  \bibfield  {author} {\bibinfo {author} {\bibfnamefont {D.}~\bibnamefont
  {Pekin}}, \bibinfo {author} {\bibfnamefont {Y.}~\bibnamefont {Skhiri}},
  \bibinfo {author} {\bibfnamefont {J.-C.}\ \bibnamefont {Baret}}, \bibinfo
  {author} {\bibfnamefont {D.}~\bibnamefont {{Le Corre}}}, \bibinfo {author}
  {\bibfnamefont {L.}~\bibnamefont {Mazutis}}, \bibinfo {author} {\bibfnamefont
  {C.}~\bibnamefont {{Ben Salem}}}, \bibinfo {author} {\bibfnamefont
  {F.}~\bibnamefont {Millot}}, \bibinfo {author} {\bibfnamefont
  {A.}~\bibnamefont {{El Harrak}}}, \bibinfo {author} {\bibfnamefont {J.~B.}\
  \bibnamefont {Hutchison}}, \bibinfo {author} {\bibfnamefont {J.~W.}\
  \bibnamefont {Larson}}, \bibinfo {author} {\bibfnamefont {D.~R.}\
  \bibnamefont {Link}}, \bibinfo {author} {\bibfnamefont {P.}~\bibnamefont
  {Laurent-Puig}}, \bibinfo {author} {\bibfnamefont {A.~D.}\ \bibnamefont
  {Griffiths}}, \ and\ \bibinfo {author} {\bibfnamefont {V.}~\bibnamefont
  {Taly}},\ }\href {\doibase 10.1039/c1lc20128j} {\bibfield  {journal}
  {\bibinfo  {journal} {Lab Chip}\ }\textbf {\bibinfo {volume} {11}},\ \bibinfo
  {pages} {2156} (\bibinfo {year} {2011})}\BibitemShut {NoStop}%
\bibitem [{\citenamefont {Pompano}\ \emph {et~al.}(2011)\citenamefont
  {Pompano}, \citenamefont {Liu}, \citenamefont {Du},\ and\ \citenamefont
  {Ismagilov}}]{Pompano2011}%
  \BibitemOpen
  \bibfield  {author} {\bibinfo {author} {\bibfnamefont {R.~R.}\ \bibnamefont
  {Pompano}}, \bibinfo {author} {\bibfnamefont {W.}~\bibnamefont {Liu}},
  \bibinfo {author} {\bibfnamefont {W.}~\bibnamefont {Du}}, \ and\ \bibinfo
  {author} {\bibfnamefont {R.~F.}\ \bibnamefont {Ismagilov}},\ }\href {\doibase
  10.1146/annurev.anchem.012809.102303} {\bibfield  {journal} {\bibinfo
  {journal} {Annu. Rev. Anal. Chem.}\ }\textbf {\bibinfo {volume} {4}},\
  \bibinfo {pages} {59} (\bibinfo {year} {2011})}\BibitemShut {NoStop}%
\end{thebibliography}%

\end{document}